\documentclass[pre,showpacs,amsmath]{revtex4}
\usepackage{amsmath,bm}
\usepackage{epsfig}
\usepackage{color,graphicx}
\begin{document}

\begin{center}
{\LARGE {Phase disorder on laser beam combining in a Talbot resonator}} \\[0pt]
\vspace*{1cm}

{\large Yeojin Chung}, {\large Alejandro B. Aceves}\\[0pt]
Department of Mathematics, Southern Methodist University,
Dallas, TX 75275, USA.\\[0pt]
\end{center}
\vspace*{2cm}

We study the role that phase disorder plays in the mode discrimination characteristics of a
Talbot resonator. Two cases considered here correspond to a six-core and a twelve-core 
photonic crystal fiber.   
\vspace*{2cm}

\section{Introduction}
Fiber laser arrays are currently considered as one of the most viable options for
state of the art high power lasers \cite{richard}. Key to their effectiveness is to coherently
combine the output of as large a number of amplifiers as possible. Coherent beam
combining can be implemented either passively \cite{huo04,hans,huo,bochove,liu,shirakawa,wu}, or 
actively \cite{shay}, with the first one being preferred
given that in principle it would be easier to implement. Passive schemes are  based on
linear mode coupling of optical elements and with the advancement of technology in multicore fiber arrays, 
many of the recent schemes make use of such arrays. Regardless of the beam-combining architecture
in all instances, we define $E_{output}= \sum_i\sum_j {\bf S}_{ij}E_j$ as the output field that results from 
combining {\it N} electric fields $E_j$ emerging from individual and ideally identical amplifiers. 
Here ${\bf S}_{ij}$  defines the passive coupling scheme. The objective of any combining scheme is 
effectively  one of efficient mode selectivity or mode discrimination. This is the case if for example 
there is an  eigenmode of ${\bf S}_{ij}$ whose (largest) eigenvalue is much bigger than the rest. Furthermore
maximum coherence is achieved if the dominant mode is such that the phases of 
$E_i^{(out)}= \sum_j {\bf S}_{ij}E_j$ are identical. Having said this, it is typical that
in most if not all instances, as the number of elements $N$ in the arrays increase, the coherence 
decreases \cite{kouznetsov}. This is due to the fact that most of perturbations 
appearing during the propagation of the fields in the (ideally identical) individual amplifiers affect the 
path length, thus in turn the phase. This combined with the fact that as $N$ increases mode discrimination
decreases, one expects coherence to diminish. For each passive linear scheme, it is important
to characterize its robustness to phase disorder; in particular it  would be useful to determine
how the coherence changes when there is disorder in the input phases and how it scales with
$N$. 

Given the fast technological advances in multicore fibers, they present perhaps the most suitable 
architecture for passive beam combining and in this paper we discuss such multicore arrays in a Talbot 
resonator \cite{talbot} as proposed in \cite{mafi}. The natural modes in this resonator are due to self-imaging, 
which according to the Talbot effect happens to images formed by periodic objects, such as a 
periodic diffraction  grating. The combination of periodicity and Fresnel diffraction is such that the
image (mode) is reproduced every multiple of the Talbot length $z_T= \frac{2a}{\lambda}$, 
where $a$ is the period of the grating and $\lambda$ is the light wavelength. 

The next section presents the main results highlighting the role of phase disorder on mode selectivity. 
While our work deals with one particular beam combining scheme, we believe this behavior is generic of
all {\it passive, linear} schemes. This is why in section \ref{nonlinear} we briefly discuss nonlinear schemes
and their suitability to achieve high coherence for large arrays.

\section{Talbot resonator}
In this section, we present the results of numerical simulation about the role of phase disorder on beam combining. 
We limit our study to the scheme based on a Talbot resonator  
for a photonic crystal fiber geometry \cite{mafi}. We will consider two periodic like profiles: a hexagonal one 
consisting of six cores, equally spaced, one at the center of the hexagon. The second case is a rectangular 
array of 12 cores. In either case the generated modes will take the form  

\begin{equation}
E_m(x,y,0)=\sum_{j=1}^n \exp(-a[(x-x_j)^2+(y-y_j)^2])e^{i\phi_j^{(m)}}
\end{equation}
with $n=6$ or $n=12$.

The figure of merit to be studied is the correlation function,

\begin{equation}
\gamma(z)= \frac{|\int E^*_m(x,y,z=0)E_m(x,y,z) dxdy|}{\int|E_m(x,y,z=0)|^2 dxdy},
\end{equation}

\noindent
where ideally you would like to work in a regime for which $\gamma$ is not only high for the
in-phase mode, but separates well from other modes. These two properties
are, for the Talbot resonator what determines if mode selectivity is achieved.

\begin{figure}
\includegraphics[width = 3.4 in]{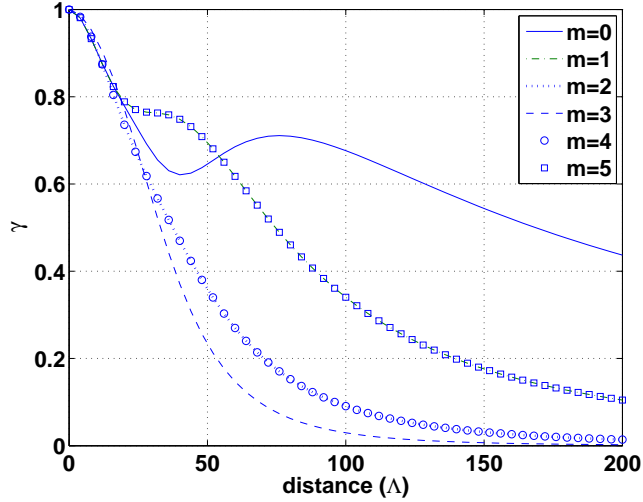}
\caption{Dependence of correlation function $\gamma(z)$ on distance from the output of waveguide 
in a 6-core PCF. The phase factor $\phi_j^{(m)}$ is considered deterministic. The distance $z$ is 
in units of lattice constant $\Gamma$, where $\Lambda/\lambda = 4.51$.}             
\label{gamma_6n}
\end{figure}

\begin{figure}
\includegraphics[width = 3.4 in]{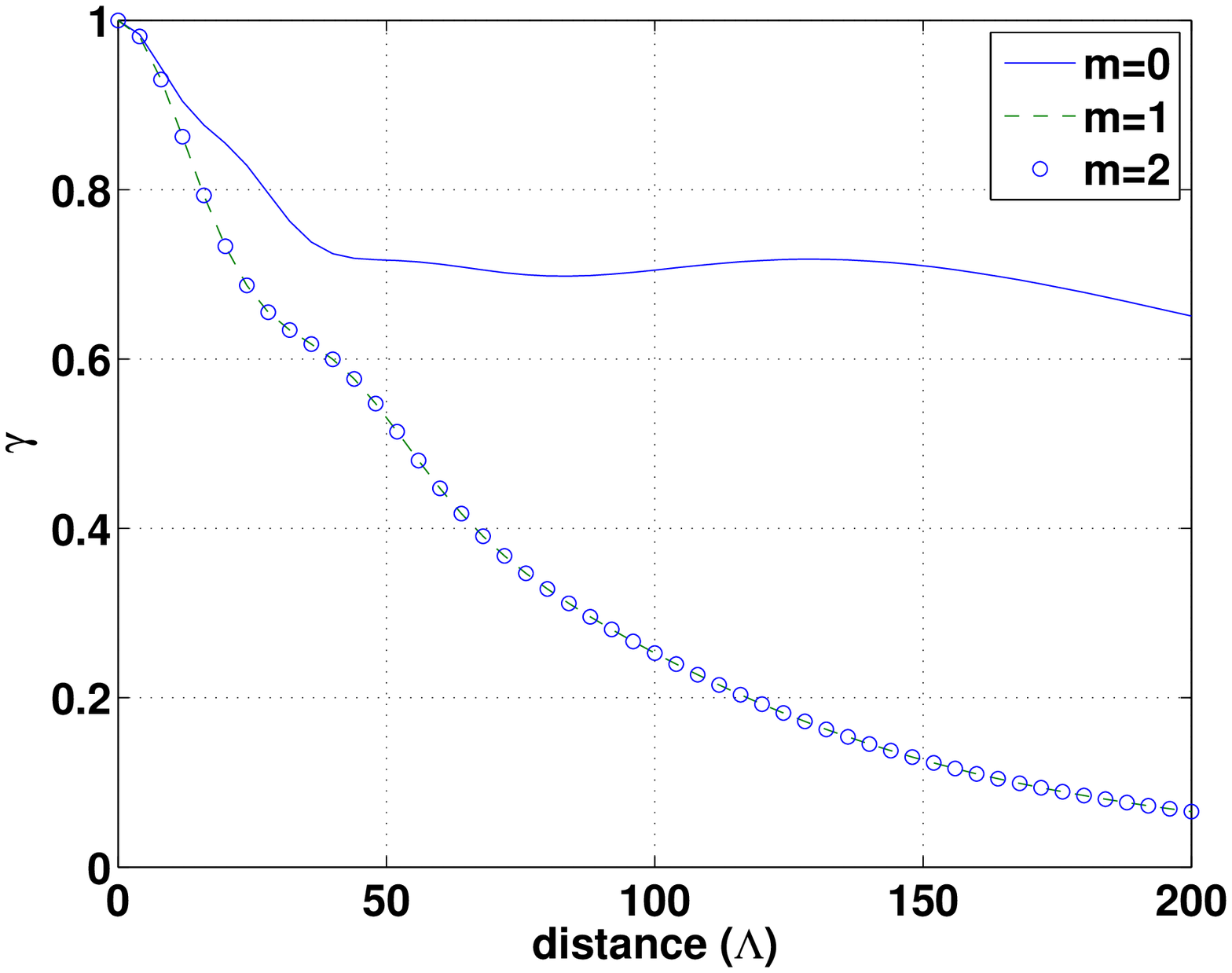}
\caption{Dependence of correlation function $\gamma(z)$ on distance from the output of waveguide 
in a 12-core PCF. The phase factor $\phi_j^{(m)}$ is considered deterministic. The distance $z$ is 
in units of lattice constant $\Gamma$, where $\Lambda/\lambda = 4.51$.} 
\label{gamma_12n}
\end{figure}

Figures \ref{gamma_6n},\ref{gamma_12n} show the dependence of correlation function $\gamma(z)$ 
on distance from the output of waveguide in a 6-core and 12-core PCFs, respectively. 
The distance is in units of the lattice constant 
$\Lambda$, where $\Lambda/\lambda=4.51$. For each supermode $m$, we consider the deterministic 
phase factor $\phi_j^{(m)} = 2m(j-1)\pi/6$ 
associated with core $j$, where $j=1,\cdots,6$ for 6-core PCF and $j=1,\cdots,12$ for 12-core PCF. 
As shown in both figures \ref{gamma_6n},\ref{gamma_12n}, it is the case that in the absence of disorder 
both the $N=6$ cores in a hexagonal  lattice and the $N=12$ cores in a rectangular lattice, mode selectivity
is achieved. We should point out that figure \ref{gamma_6n} reproduces the figure 9 in \cite{mafi}.

\begin{figure}
\includegraphics[width = 3.4 in]{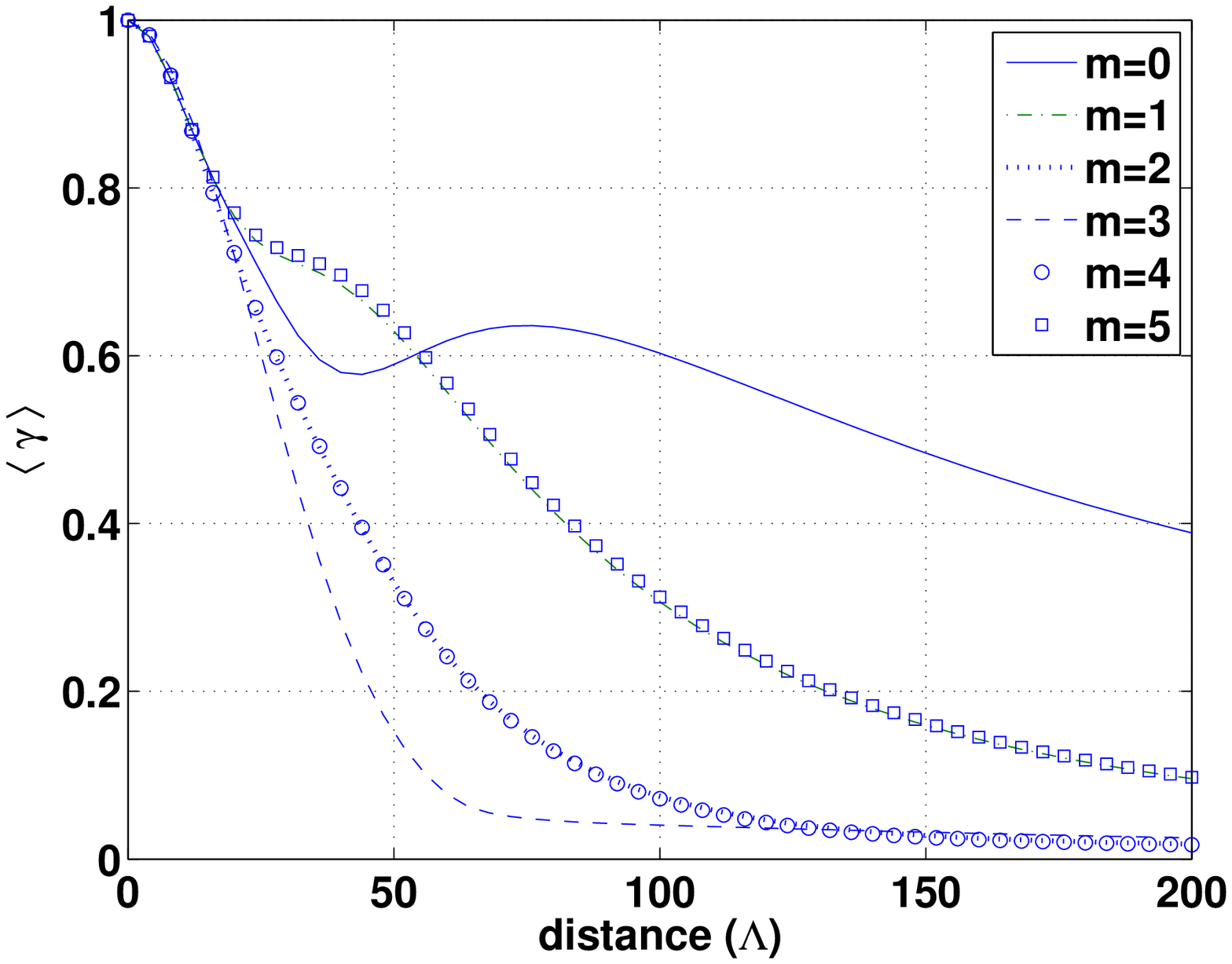}
\caption{Average of $\gamma(z)$ for $N=6$. Random phase factor is 
considered as $\phi_j^{(m)} = 2m(j-1)\pi/6+\xi$ 
for $0\leq \xi \leq \pi/2$.}             
\label{ave_6n_025r}
\end{figure}

\begin{figure}
\includegraphics[width = 3.4 in]{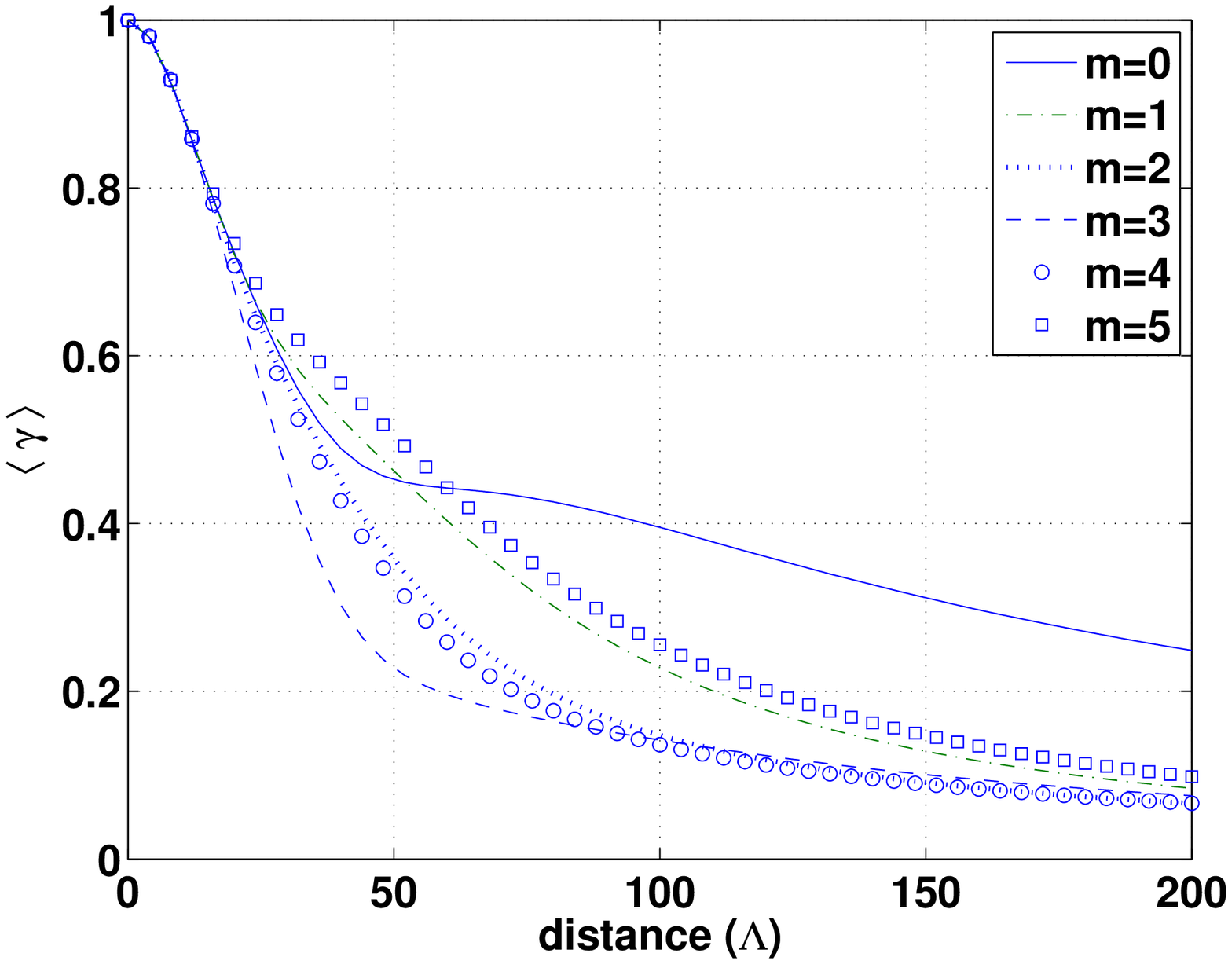}
\caption{Average of $\gamma(z)$ for $N=6$. Random phase factor is considered 
as $\phi_j^{(m)} = 2m(j-1)\pi/6+\xi$ 
for $0\leq \xi \leq \pi$.}             
\label{ave_6n_05r}
\end{figure}

\begin{figure}
\includegraphics[width = 3.4 in]{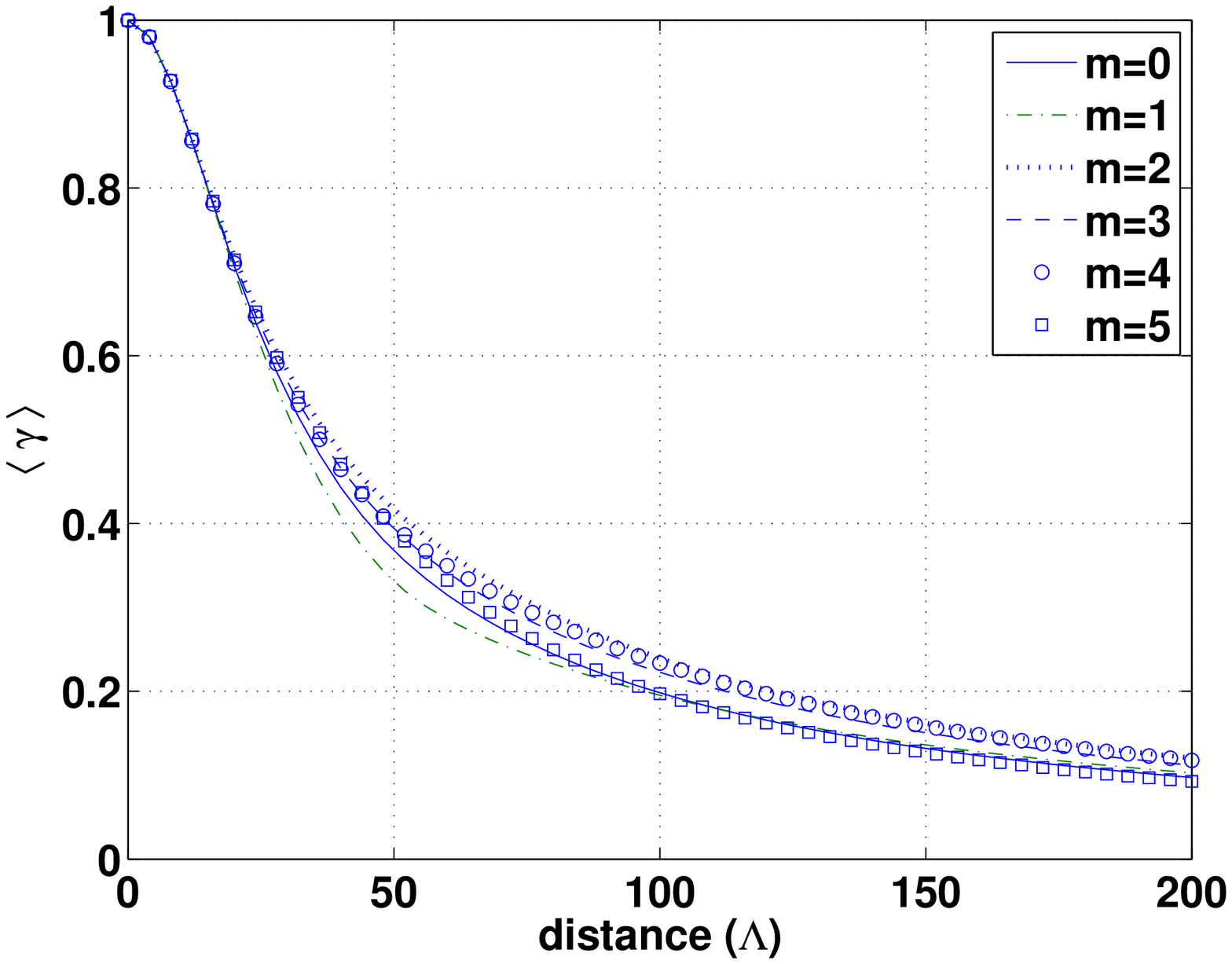}
\caption{Average of $\gamma(z)$ for $N=6$. Random phase factor is considered as 
$\phi_j^{(m)} = 2m(j-1)\pi/6+\xi$ for $0\leq \xi \leq 2\pi$.}             
\label{ave_6n_r}
\end{figure}

\begin{figure}
\includegraphics[width = 3.4 in]{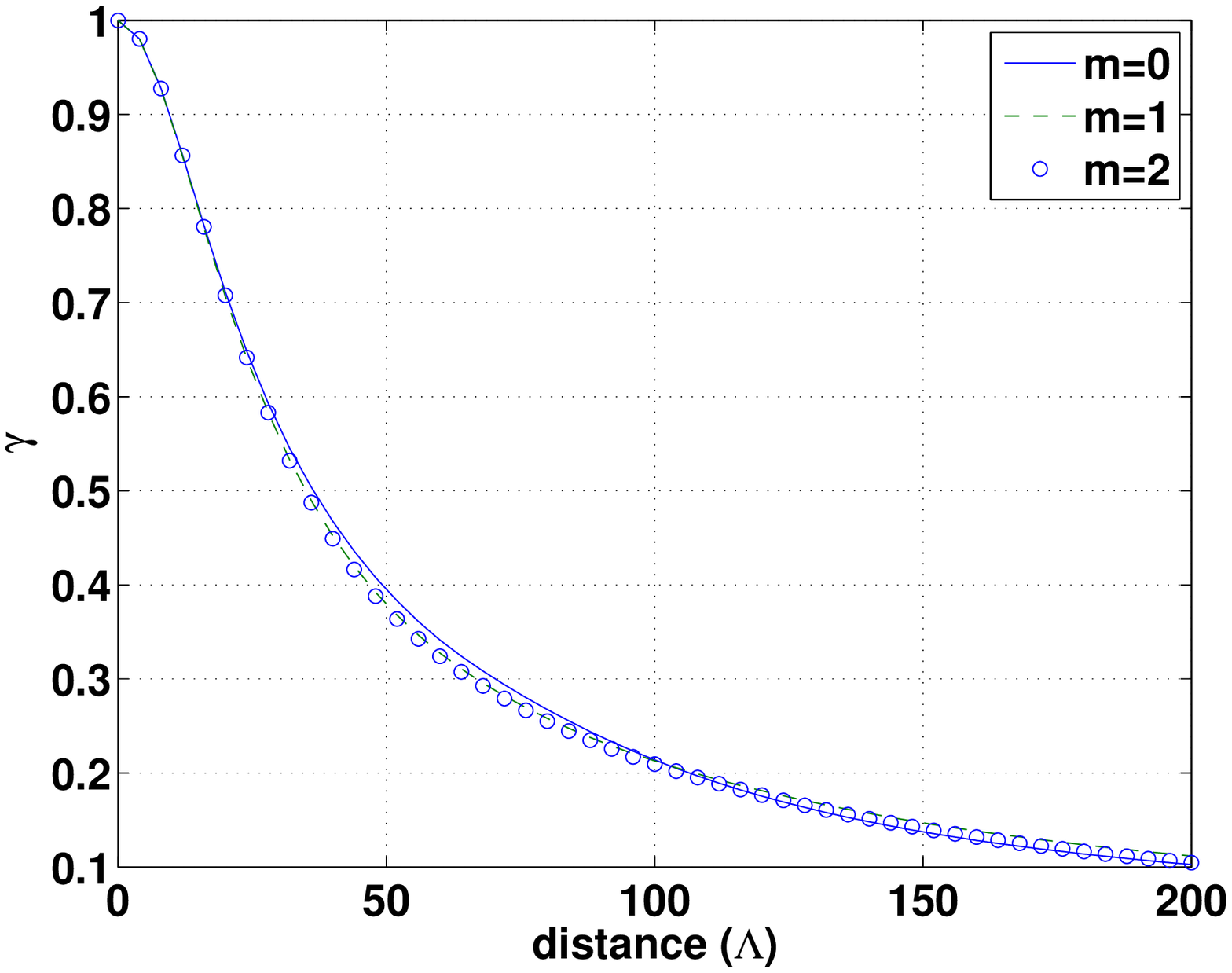}
\caption{Average of $\gamma(z)$ for $N=12$. Random phase factor is considered as 
$\phi_j^{(m)} = 2m(j-1)\pi/3+\xi$ for $0\leq \xi \leq 2\pi$.}             
\label{ave_12n_r}
\end{figure}

The rest of the figures represent averaged values $\langle\gamma \rangle$ computed over many realizations of 
different strengths of the random 
phase perturbations. We assume in all instances the phase inputs are {\it iid}. Figures 
\ref{ave_6n_025r}, \ref{ave_6n_05r}, \ref{ave_6n_r} correspond to different realizations of the random phase 
factor in a 6-core PCF with 
same parameters as figure \ref{gamma_6n} for increasing disorder strengths. That is, in our simulations the phase 
includes a random variable $\xi$
uniformly distributed so that  $\phi_j^{(m)} = 2m(j-1)\pi/6 + \xi$ and with strengths 
$0\leq \xi \leq \pi/2$ (figure \ref{ave_6n_025r}), $0\leq \xi \leq \pi$ (figure \ref{ave_6n_05r}) and 
$0\leq \xi \leq 2\pi$ (figure \ref{ave_6n_r}). We clearly observe that while the  contrast (separation of 
different curves) between the in-phase  ($m=0$) mode and the others does not diminish by much for
noise levels  $0\leq \xi \leq \pi/2$ (figure 3) it does so in a significant way once the noise strength 
increases (figure 4). That is, 
even if one were  able to control path lengths so that the noise is only half a period $0\leq \xi \leq \pi$ 
this scheme becomes ineffective. Notice that if  $0\leq \xi \leq 2\pi$ (figure \ref{ave_6n_r}) there is
no mode discrimination at all. We should notice that for the operational wavelengths 
in fiber lasers, small perturbations (e.g., nonlinear phase, fiber lengths, thermal effects, gain induced 
perturbations to name some ) are likely 
to modify the phase in multiples of $2\pi$. Figure \ref{ave_12n_r} shows the average 
of $\gamma(z)$ over different realizations of random phase factor in a 12-core PCF, 
where $\phi_j^{(m)} = 2m(j-1)\pi/3 + \xi$ for $j=1,\cdots,12$, $m=0,1,2$, and $0\leq \xi \leq 2\pi$. The 
similarity of the outcome  with the 6-core case suggests that noise will ``erase'' the contrast irrespective of 
the number of fibers or the geometry.

\section{Nonlinear coupling schemes} \label{nonlinear}
The previous section highlights the sensitivity to phase disorder and the loss of coherent beam combining 
efficiency. Furthermore, it is generally the case (with or without disorder) that as the number of elements 
in a fiber laser array increases, the coherence diminishes for linearly based coupling schemes \cite{kouznetsov}. The role of phase disorder in linearly based coupling schemes we believe will be similar to what we just 
showed forthe Talbot resonator discussed above. Altogether, 
new paradigms need to be sought. Here we
suggest that having a nonlinear coupling scheme should be considered as a viable alternative. As we know
from many studies not only in laser dynamics but in other areas of science, synchronization of weakly
coupled oscillators is universal. Whether one describes laser modes, social networks, fireflies or chemical 
reactions to name some, a phase transition where incoherent oscillations lock into an in-phase state
is present at a critical coupling strength. For this to happen, the system requires to be nonlinear,
but the details need not be identical. This behavior is best represented in the work by Kuramoto
on a ``globally'' weakly coupled nonlinear chain of oscillators,

\begin{equation}
\frac{d\theta_i}{dt}={\tilde \omega_i} +\frac{K}{N}\sum_{j=1}^N\sin(\theta_i-\theta_j), \ \ i=1, 2,..., N,
\end{equation}

\noindent
where $\theta_i(t)$ is the phase of the $i-th$ oscillator and ${\tilde \omega_i}$ is its natural 
frequency. In this model, disorder comes from the fact that the frequencies are random with a Lorenzian 
probability density

\begin{equation}
g(\omega)= \frac{\delta}{\pi [\delta^2+(\omega-\omega_0)^2]}.
\end{equation}

Kuramoto \cite{kuramoto} found a phase transition in that the order parameter

\begin{equation}
r(t)=| \frac{1}{N}\sum_{j=1}^N e^{i\theta_j(t)} |
\end{equation}

\noindent
takes the asymptotic ($N\rightarrow\infty$ and $t\rightarrow\infty$) value

\begin{eqnarray}
r & =& 0, \;\;\; \;\;\;\;\;\;\;\;\;\;\;\;\;K < 2\delta   \\
 & =& \sqrt{1-\frac{2\delta}{K}}, \;\;\; K\ge2\delta  
\end{eqnarray}

\noindent
which translates into enhanced locking as the coupling strength goes past a critical value.

Perhaps more relevant case to coherent beam combining which can still be viewed as some form of phase 
transition is that of spatial and spatio-temporal localization in fiber arrays. Here the 
nonlinear index of refraction together with linear nearest neighbor coupling is modeled by
the discrete nonlinear Scr\"odinger equation (DNLSE), 

\begin{equation}
i\frac{\partial a_{nm}}{\partial z}= (\Delta{\bf\underline a})_{nm} + n_2|a_nm|^2a_{nm},
\end{equation}

\noindent 
where the term $\Delta {\bf \underline a}$ represents a general linear coupling scheme.
Specific to cw lasers, the possibility of reaching an output that is spatially localized into
a few cores would enhance coherence if it is measured by

\begin{equation}
r=\frac{|\sum_{nm} a_{nm}|^2}{\sum_{nm} |a_{nm}|^2}.
\end{equation}

\begin{figure}
\includegraphics[width = 3.4 in]{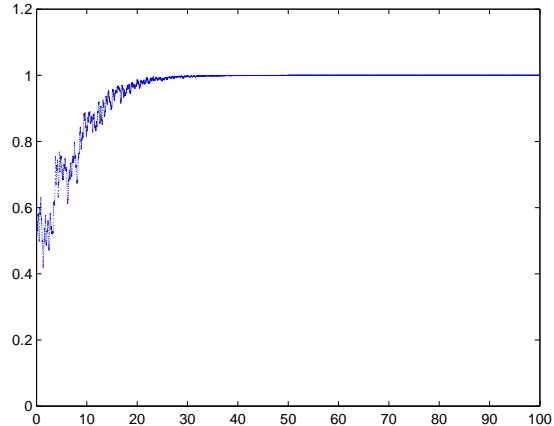}
\caption{Time evolution of  $E= \frac{|\sum_1^N u_n|}{\sum_1^N|u_n|}$ for an initial 
condition with  $E = 0.528646241$ for an array of 17 nonlinear coupled oscillators model \cite{panos}}             
\label{pla2d}
\end{figure}

As theory \cite{aceves} and experiment \cite{minardi} have shown, at sufficiently high powers 
(the equivalent of having high enough $K$ in the oscillator model), an initially broad distribution 
of power reaches in propagation a localized mode. We are currently exploring the role that the disorder 
of incident phases plays in the dynamics of spatial and spatio-temporal localization. It is also
the case that non Hamiltonian perturbations to coupled systems like the DNLSE may produce an attractor in
the form of a highly coherent state. As a proof of principle, recent work \cite{panos} on large systems
of nonlinearly coupled oscillators shows one can find asymptotic dynamics towards an attracting highly 
coherent state. Figure (\ref{pla2d}) shows a dynamics for an array of 17 oscillators reaching a state
of maximum coherence.

Finally, nonlinear coupling schemes can perform in a very robust manner even in the presence of
losses. In fact, small losses (that is departing from a conservative system where stable equilibrium
states are centers) should facilitate synchronization \cite{eckhouse}.

\section{Conclusions}
Phase disorder is always present in the amplification and propagation of light in optical 
fibers. This work presents the deterrent effect it has on the efficient beam combining in 
a Talbot resonator. Two cases of 6 and 12 elements are considered as a way to explore the
role if any,  $N$ the number of elements plays on the efficiency of coherent coupling. 
We found for these two cases that mode selectivity characterized by the separation of the parameter
$\gamma$ for different modes lost if the noise in the phases is of strength equal to 
$2\pi$. While the results are limited to one particular 
linear scheme, we believe all linear schemes will show an eventual decrease in efficiency not only
by increasing the number of elements as previously stated \cite{kouznetsov}, but also
in terms of expected phase disorder generated during propagation in the individual fiber amplifiers. 
The last section provides a  brief discussion of a new paradigm whereby considering nonlinear coupling
mechanisms, phase transitions towards high coherence are at least theoretically feasible in
the important $N\rightarrow\infty$ case.

\section{Acknowledgements}
 A. Aceves acknowledges the collaboration and fruitful discussions with Dr. Erik Bochove from 
the Air Force Research Laboratory.

\end{document}